\input harvmac
%\draftmode
\noblackbox

%
% Poor man's Blackboard Bold characters often used :
%
\def\inbar{\,\vrule height1.5ex width.4pt depth0pt}
\def\IB{\relax{\rm I\kern-.18em B}}
\def\IC{\relax\hbox{$\inbar\kern-.3em{\rm C}$}}
\def\ID{\relax{\rm I\kern-.18em D}}
\def\IE{\relax{\rm I\kern-.18em E}}
\def\IF{\relax{\rm I\kern-.18em F}}
\def\IG{\relax\hbox{$\inbar\kern-.3em{\rm G}$}}
\def\IH{\relax{\rm I\kern-.18em H}}
\def\II{\relax{\rm I\kern-.18em I}}
\def\IK{\relax{\rm I\kern-.18em K}}
\def\IL{\relax{\rm I\kern-.18em L}}
\def\IM{\relax{\rm I\kern-.18em M}}
\def\IN{\relax{\rm I\kern-.18em N}}
\def\IO{\relax\hbox{$\inbar\kern-.3em{\rm O}$}}
\def\IP{\relax{\rm I\kern-.18em P}}
\def\IQ{\relax\hbox{$\inbar\kern-.3em{\rm Q}$}}
\def\IR{\relax{\rm I\kern-.18em R}}
\font\cmss=cmss10 \font\cmsss=cmss10 at 7pt
\def\IZ{\relax\ifmmode\mathchoice
{\hbox{\cmss Z\kern-.4em Z}}{\hbox{\cmss Z\kern-.4em Z}}
{\lower.9pt\hbox{\cmsss Z\kern-.4em Z}}
{\lower1.2pt\hbox{\cmsss Z\kern-.4em Z}}\else{\cmss Z\kern-.4em Z}\fi}

%
% definitions
%
\def\half{{1\over2}}

\def\({\left (}
\def\){\right )}
\def\halfsq{{1\over \sqrt{2}}}

%
% Journals
%

\def\NP{{\it Nucl. Phys.\ }}

\def\PL{{\it Phys. Lett.\ }}
\def\PR{{\it Phys. Rev.\ }}
\def\PRL{{\it Phys. Rev. Lett.\ }}

\def\Mod{{\it Mod. Phys. Lett.\ }}

%
%Fonts
%
\font\tiau=cmcsc10

%
%References
%
\lref\jp{J. Polchinski, hep-th/9510017, \PRL {\bf 75} (1995) 4274.}
\lref\vafwit{C. Vafa and E. Witten, hep-th/9507050.}
\lref\sqms{S. Ferrara, J.A. Harvey, A. Strominger and C. Vafa, hep-th/9505162,
\PL {\bf B361} (1995) 59.}
\lref\ms{J. Maldacena and A. Strominger, hep-th/9603060, \PRL {\bf 77} (1996)
428.}
\lref\cliffordfd{ C. Johnson,
R. Khuri, R. Myers, hep-th/9603061, \PL {\bf B378} (1996) 78.}
\lref\ascv{A. Strominger and C. Vafa, hep-th/9601029, \PL {\bf B379} (1996)
99.}
\lref\cama{C. Callan and J. Maldacena, hep-th/9602043, \NP {\bf B472}
 (1996) 591.}
\lref\kol{R. Kallosh and B. Kol, hep-th/9602014, \PR {\bf D53} (1996) 5344; 
R. Dijkgraaf,
E. Verlinde and H. Verlinde, hep-th/9603126.}
\lref\tf{A. Tseytlin, hep-th/9601177, \Mod {\bf A11} (1996) 689.}
\lref\openstrom{A. Strominger, hep-th/9512059, \PL {\bf B383} (1996) 44.}
\lref\opentown{P. Townsend, hep-th/9512062, \PL {\bf B373} (1996) 68.}
\lref\jmphd{ J. Maldacena, hep-th/9607235.}
\lref\spin{ J. Breckenridge, R. Myers, A. Peet and C. Vafa, hep-th/9602065. }
\lref\wittenpol{ J. Polchinski and E. Witten, hep-th/9510169, 
\NP {\bf B460} (1996) 506.} 
\lref\dougauge{M. Douglas, hep-th/9604198.}
\lref\msu{ J. Maldacena and L. Susskind, hep-th/9604042,
\NP {\bf B475}(1996) 679.}
\lref\classfour{ M. Cvetic and D. Youm, hep-th/9512127; M. Cvetic
and A. Tseytlin, hep-th/9512031, \PR {\bf D53} (1996) 5619.}
\lref\gimon{E. Gimon and J. Polchinski, hep-th/9601038, 
\PR {\bf D54} (1996) 1667.}
\lref\fks{S. Ferrara, R. Kallosh and A. Strominger, hep-th/9508072,
\PR {\bf D52} (1995) 5412.}
\lref\as{A. Strominger, hep-th/9602111, \PL {\bf B383} (1996) 39.} 
\lref\fk{S. Ferrara and R. Kallosh,
hep-th/9602136, \PR {\bf D54} (1996) 1514; 
hep-th/9603090, \PR {\bf D54} (1996) 1525.}
\lref\dmw{M.J. Duff, R. Minasian and E. Witten, hep-th/9601036, \NP
{\bf 465} (1996) 413.}
\lref\maharana{J. Maharana and J.H. Schwarz, hep-th/9207016, \NP
{\bf 390} (1993) 3.}
\lref\seiwit{N. Seiberg and E. Witten, hep-th/9603003,
 \NP {\bf B471} (1996) 121.}
\lref\berk{M. Berkooz, R.G. Leigh, J. Polchinski, J.H. Schwarz, N. Seiberg
and E. Witten, hep-th/9605184.}

%
%Title page
%
\baselineskip 12pt
\Title{\vbox{\baselineskip12pt 
\line{\hfil  CALT-68-2076}
\line{\hfil  RU-96-88}
\line{\hfil \tt hep-th/9609204}}}
{\vbox{\hbox{\centerline{Microscopic Entropy of N=2 Extremal 
Black Holes}}}}
\centerline{\tiau David
M. Kaplan$^\dagger$\foot{dmk@cosmic1.physics.ucsb.edu},
 David A. Lowe$^\sharp$\foot{lowe@theory.caltech.edu}, 
Juan M. Maldacena$^\ddagger$\foot{malda@physics.rutgers.edu} 
and Andrew Strominger$^\dagger$\foot{andy@denali.physics.ucsb.edu}}
\vskip .1in
\centerline{$^{\dagger}${\it Department of Physics}}
\centerline{\it University of California}
\centerline{\it Santa Barbara, CA 93106-9530, USA}
\vskip .1in
\centerline{$^\sharp${\it California Institute of Technology}}
\centerline{\it Pasadena, CA 91125}
\vskip.1in
\centerline{$^{\ddagger}${\it  Department of Physics and Astronomy }}
\centerline{\it Rutgers University}
\centerline{\it Piscataway, NJ 08855, USA}
\vskip .5cm
\centerline{\bf Abstract}
\noindent
String theory is used to compute 
the microscopic entropy for several examples 
of black holes in compactifications with $N=2$ 
supersymmetry.  Agreement with the 
Bekenstein-Hawking entropy and the moduli-independent  $N=2$ 
area formula is found in all cases.

\Date{September, 1996}

\newsec{Introduction}

More than two decades after its discovery, our understanding of 
string theory has finally 
developed to the point where it can be used to provide, in special
cases, a precise statistical derivation of the thermodynamic 
Bekenstein-Hawking
area law for the entropy. While some definite 
relations between the laws of black hole thermodynamics and the
statistical 
mechanics of stringy microstates have been clearly established, much 
more remains to be understood. For example a universal derivation 
of the area law for all types of black holes remains elusive. 
For these reasons it is important to understand as many cases as 
possible.

In \ascv\ the entropy was microscopically computed for the simplest 
case of a five-dimensional extremal black hole. 
Supersymmetry makes it possible to count the microstates at weak coupling
and then extrapolate into the strongly-coupled black hole region.
This result was extended to
include 
rotation in \spin. The four dimensional case with $N=8$ and $N=4$ 
supersymmetry was microscopically computed in \ms\ and \cliffordfd.
The result agreed with the formulae for the area derived 
in \refs{\classfour,  \kol}. Given the fact that 
the area is independent of moduli \refs{\fks, \classfour}, 
these formulae are fixed up to a few 
constants by symmetries of $N=8$ and
$N=4$ supergravity. In this paper we analyze several examples with 
$N=2$ supersymmetry, which has not been previously 
considered. The $N=2$ area formula was derived for the pure 
electric case in \fks, for the general case with electric and
magnetic charges in \as\ and elegantly related to central charge
minima 
in \fk.  No symmetries are available here, and 
the formulae have a rather different character involving 
rational fixed points in the special geometry moduli space. 
Agreement is again found in the examples considered herein.    

\newsec{Type IIA Orientifold Example}

First we consider the Type IIA theory on $K3\times S^1\times \hat S^1$.  This
theory has $N=4$ supersymmetry in $d=4$.  In order to construct an $N=2$
theory, we orientifold this model by a geometric $\IZ_2$ 
symmetry  combined with reversal of
worldsheet orientation.  Black hole solutions of the $N=4$ theory 
invariant under this action will also be solutions of the orientifold model.  
In this manner we construct black hole solutions of an $N=2$ Type IIA 
orientifold model and compute their macroscopic and microscopic entropy.

We consider the $\IZ_2$ orientifold which combines the
Enriques involution on $K3$, reflection on $S^1$, and translation by
$\pi$ on $\hat S^1$, together with left-right exchange on the
worldsheet. In M-theory language, this is a purely geometric orbifold 
of $K3\times S^1\times
\hat S^1\times S^1_{11}$, which acts as Enriques, combined with
reflection on $S^1\times S^1_{11}$ and translation by $\pi$ on $\hat
S^1$.  This model was discussed in a different construction 
in \sqms\ (with the translation on $S^1_{11}$ rather than $\hat
S^1$) and with this construction in \vafwit. It has N=2 supersymmetry 
in d=4 with 11 vector multiplets
and 12 hypermultiplets.  

Now we wish to construct a four-dimensional black hole solution in this 
$N=2$ theory as a collection of intersecting branes \jp. Such a solution
is obtained from a set of intersecting branes in the original $N=4$
theory as follows.
First, $Q_5$ symmetric 5-branes wrap
$K3\times\hat S^1$. We will consider 5-branes
not centered at the fixed points of the reflection
on $S^1$. Since we wish to construct
a configuration invariant under the $\IZ_2$
symmetry each 5-brane is accompanied by its
$\IZ_2$ image, so that $Q_5$ is even.  
In addition, a 4-brane wraps
the product of a holomorphic 2-cycle, $\Sigma$, of $K3$ with
$S^1\times \hat S^1$. It is possible to show that for any $\Sigma\in H_*(K3)$
with even self-intersection number $m=2Q_4^2$, there is a choice of complex
structure of $K3$ such that the cycle may be realized as a holomorphic
curve of genus $Q_4^2 +1$. Since $S^1\times \hat S^1$ is odd under
the $\IZ_2$, we must impose the additional restriction that
the 2-cycle of $K3$ be odd under the
Enriques involution. In order that the 4-cycle be supersymmetric in the
orientifold theory, the 2-cycle must be holomorphic with respect
to the odd self-dual two-form of $K3$. Note also that the symmetric
5-branes cut each of the 4-cycles into $Q_5$ pieces along the
$S^1$ direction. 
Finally, we include $n$ quanta of momentum along the $\hat S^1$ direction.

In terms of the original $N=4$ theory, we are considering a set
of $Q_5$ symmetric 5-branes,  a 4-brane with self-intersection
number $2Q_4^2$ and total momentum $n$. 
This configuration is related by duality 
to a configuration of one 6-brane and
$Q_4^2+1$ 2-branes, $Q_5$ 5-branes and momentum
$n$.  The statistical entropy of
such a configuration was calculated in \ms\ and
found to be 
\eqn\macentrop{
S= 2 \pi \sqrt{Q_4^2 Q_5 n}~,
}
in the limit of large charges. This agrees with the 
Bekenstein-Hawking entropy of the corresponding black hole solution. 
The entropy is duality invariant, 
so \macentrop\ will also hold in the case at hand.

Now let us consider the Bekenstein-Hawking entropy
in the orientifold theory. 
This may be computed by dividing the horizon area $A_{10}$ of the 
ten-dimensional solution by the ten-dimensional Newton's constant
$G_{10}$. 
Newton's constant is unaffected by the orientifolding but 
the $Z_2$ acts freely on the horizon and hence divides the area in 
half. The entropy $S^\prime$ in the orientifold theory is then 
\eqn\ssp{S^\prime={A_{10}^\prime \over 4 G_{10}^\prime}=
{A_{10} \over 8 G_{10}}={S \over 2},}
where the prime denotes quantities in the orientifold theory. 

It is also easily seen that 
orientifolding reduces the microscopic entropy by half. 
The microscopic entropy is carried by $Q_4^2 Q_5$ massless
supermultiplets that live on a string wrapping  $\hat S^1$.
The orientifold reduces the length of this string by half. 
Since the entropy is an extensive quantity it is also reduced 
by half. The orientifold also introduces twisted spatial 
boundary 
conditions for the massless supermultiplets. However this affects 
mainly the zero mode structure and not the asymptotic form of the
entropy for large $n$.

\newsec{Type I Example}

In this section 
we consider black holes in 
Type I theory on K3.  These theories have $N=2$ supersymmetry in
four dimensions when we further compactify two more dimensions on a torus. 
First we describe 
the classical black hole solutions, then
quantize the charges in four and five dimensions and
compute the Bekenstein-Hawking entropy. This is found to agree with the 
number of microscopic configurations  obtained using D-brane techniques.

\subsec{Classical Solutions}

For the five(four)-dimensional black holes we
consider Type I on K3$\times S^1$($\times \hat S^1$).
The four-dimensional classical solution was found in \classfour ~for the 
case of toroidal compactification.  The solution in the $N=2$
case is the same as the $N=4$ case as the relevant terms in the low
energy supergravity lagrangians involved are identical. In four dimensions
the black hole we consider carries charges corresponding to 
NS solitonic 5-branes wrapping
around K3$\times S^1$, Kaluza-Klein  monopoles on $\hat S^1$, 
and 
fundamental string winding and momentum along $S^1$.
The five-dimensional configuration is the same, except for the absence of
the Kaluza-Klein  monopole. The five dimensional solutions are treated
in \tf.

It is useful to rewrite the entropy formulas in terms of integer
quantized charges.
The fundamental strings are winding along $S^1$ so the charge
quantization
condition will be the same as in $N=4$. The quantization for 
momentum will  be the same. The NS 5-brane is the Dirac dual 
to the string winding along $\hat S^1$, so it also has the same
quantum of charge as in  $N=4$ case, and the Kaluza-Klein monopole is the 
Dirac dual to momentum along $\hat S^1$ so again it is the same as
in  the $N=4$ case. 
Therefore the formula for the entropy is 
\eqn\entroclass{
S = 2 \pi \sqrt{ Q_5 Q_{KK} Q_1 n}~,
}
where $Q_5,Q_{KK}, Q_1, n$ are the number of NS 5-branes,
 Kaluza-Klein monopoles, winding strings, and momentum, respectively.
The entropy of the  five-dimensional black hole  is as in \entroclass\
with the factor $Q_{KK}$ set to one.

\subsec{D-Brane Counting}

Consider first the five-dimensional case. The Type I 
configuration on K3$\times S^1$ consists of D 5-branes
wrapping K3$\times S^1$, D-strings wrapping $S^1$ and momentum flowing
along $S^1$. Note that these are D-branes of Type I theory so that the
D 5-brane has an $SU(2)=Sp(1)$ gauge field living on it.
When $Q_5$ D-5-branes coincide we have an $Sp(Q_5)$ gauge theory on 
the brane. The D-1-brane charge is carried by instantons of this
gauge theory which are self dual gauge connections on the $K3$. 
For large $Q_1$ and $Q_5$ the number of bosonic degrees of
freedom of the instanton moduli space is $4 Q_1 Q_5$ and they come
mainly from the different ways of orienting the instantons inside the
gauge group. We could also,  as in \cama\ , count the moduli by
considering open strings going between the D 1-branes and the
D 5-branes. These open strings are unoriented, so there are
2 bosonic ground states for each string, plus 2 possible Chan
Paton factors for each 5-brane. 
This leads to a total of $4 Q_1 Q_5 $ bosonic states\foot{
The actual state will also have some (1,1) and (5,5) strings excited
in such a way that D-terms vanish \refs{\jmphd, \dougauge}.}.  (In the 
corresponding Type II
counting there is a  factor of 2 arising from the 2 possible orientations
of the open string).  There are also $4Q_1Q_5 $ fermionic
degrees of freedom. The momentum $n$ along the $S_1$ direction will be
carried 
by oscillations in the instanton moduli space or, equivalently,
by the (1,5) strings. In either case the counting is that of
a 1+1-dimensional gas with $4Q_1Q_5$ bosonic and fermionic
flavors\foot{
It should be kept in mind that when $n$ is not much bigger than
$Q_1Q_5$ then the effects of multiple windings become important and
both the number of flavors and the effective length of the circle 
increase, giving the same result for the entropy \msu .}.
We can therefore use the  
standard asymptotic formula for the entropy which yields
\eqn\entrofive{
S = 2 \pi \sqrt{ Q_1 Q_5  n}~.
} 
This agrees with the classical result.

Now let us consider the four-dimensional case. 
We have the same configuration of branes as in five dimensions
plus a  Kaluza-Klein monopole. 
It was shown in \cliffordfd\ for $N=4$ (but the arguments also apply
to 
$N=2$) for  the case of one  
monopole that \entrofive\ is unchanged, in agreement with 
\entroclass\ for $Q_{KK}=1$. The general formula then follows from 
duality. It is nevertheless instructive, as well as useful for 
following sections, to see how this works in detail.  
We first perform a T-duality along $\hat S^1$ to a
Type $I^\prime$ theory. The D 5-branes become D 6-branes and the 
D-strings become D 2-branes wrapped around  $ S^1 \times \hat S^1 $.
The momentum remains momentum and the Kaluza-Klein  monopole becomes a 
NS 5-brane wrapping K3$\times S^1$. 
This theory also has two orientifold planes
perpendicular to $\hat S^1$ and 16 D-8-branes with 16 images on the other
side of the orientifold planes.
Locally, there is a Type II theory in between the 
orientifold planes.
We take the NS 5-branes to sit at particular 
points of $\hat S^1 $. The total number of Type II 5-branes is
$2 Q_{KK}$: a 5-brane and an image 5-brane
is the minimum that we can have, therefore it is
the ``unit'' of solitonic 
5-brane charge in this theory. However only $Q_{KK} $ 
of them are in between a pair of orientifold hyperplanes so that the 
effective number of 5-branes in the locally Type II theory between 
two orientifold hyperplanes is
$Q_5^{II} = Q_{KK} $.
The number of D-2-branes in the locally Type II
theory is $Q_2^{II} = Q_1$ (the same as the number of original one
branes), while the number of locally Type II D-6-branes is
$Q_6^{II} = 2 Q_5 $. The extra factor of two comes from the
extra $Sp(1)$ Chan Paton index carried by Type I 5-branes. 
We can now
use the Type II result \ms ~to count the number of
configurations
between two orientifold planes. 
The counting in \ms ~relied on the 
fact that the solitonic 5-branes slice the 2-branes in $Q_5^{II}$
pieces.  Now we must in addition consider  
the D-8-branes, which also intersect the 2-branes, further dividing up
their worldvolumes. However for large charges they will have a subleading 
effect since the number of 8-branes is much smaller than $Q_5^{II}$
in that limit ($16 \ll Q_5^{II}$).
Similarly we will not worry about possible slicing of the
6-branes by the 8-branes\foot{We also do not worry about 
(2,8) or (6,8) strings because, for large charges, there are 
not as many flavors of these as there are for (2,6) strings.}. 
The fact that 
the 2-branes can end on 5-branes \refs{\openstrom, \opentown}  
implies that different slices in
between different solitonic 5-branes 
can move independently.
The momentum is carried mainly 
by (2,6) strings living between particular 5-branes. Note however
that
the orientifold projection will correlate what happens on one side of
the orientifold hyperplane 
with what happens on the other side. In particular 
when we put a unit of momentum between two 5-branes we also have
to put a unit of momentum on the two image 5-branes on 
other side of the orientifold 
hyperplane. 
 Therefore we have only half of the total
momentum available for distributing freely in the locally Type II
theory, $n^{II} = n/2$. 
Once we have identified the correct number of degrees of freedom in
the locally Type II theory we count as in \ms ~and 
obtain 
\eqn\entrofour{
S = 2 \pi \sqrt{ Q^{II}_2  Q^{II}_6 Q^{II}_5 n^{II}}
  = 2 \pi \sqrt{ Q_1 Q_5 Q_{KK} n}~,
}
which is the same formula as in the $N=4,8$ cases \ms.  Again, this formula
agrees with the classical result of equation \entroclass\ .  

\newsec{More General N=4 Examples}

In the previous two $N=2$ examples the counting is similar 
to the maximally supersymmetric $N=8$ cases. This is 
partly because the black holes  we chose to 
analyze were  present in $N=8$ as well as $N=2$ theories.
In this section we analyze some new features that appear 
only when there is less than the maximal supersymmetry. 
In the $N=8$ case all gauge charges are part of the supergravity
multiplet, while in the 
$N=4$ case we can  have extra gauge multiplets.
In this latter, more general case, the entropy formula in four dimensions
can be written in terms of an $O(6,22)$ 
vector of magnetic charge $ \bf P $ and
an $O(6,22)$ vector of electric charges $ \bf Q $.
We consider Type I/heterotic   theory on $ T^6 $, in which case the 
electric charges are carried by the Type I D-1-brane or heterotic
fundamental string. Define $  {\bf Q} = 
(\half {\bf p}_R, \half {\bf  p}_L, \halfsq {\bf q} ) $
where $\bf p_{R,L} $ are the right and left-moving momenta of 
a heterotic string on $ T^6 $ and $ \bf q $ are the 16 U(1) 
charges of a generic compactification. 
In terms of D-branes these charges are carried by (1,9) strings. 
The black holes
we considered in the previous section had  $ p_{R,L}^5
 =  ( {n\over R} \pm Q_1R )  $ 
(with other components of ${\bf p}_{R,L}$ set to zero) and ${\bf q}
=0$; 
now let
us consider $\bf q $ different from zero. 
The magnetic charges are still  carried by the D-5-brane and the
Kaluza-Klein monopole. As in the preceding section, we go to the Type 
$I^\prime$ theory
with $Q_1$ D-2-branes, $Q_5$ D-6-branes and $ Q_{KK}$ 
solitonic 5-branes\foot{The subindex indicates what the object was
in
the original Type I theory, hopefully this will not cause confusion.}.
Now the open strings 
that carry momentum $n$ will also have to carry some charge.
The charge is carried by (2,8) strings, the D-8-branes appeared 
when we did the T-duality transformation to the Type $I^\prime$
theory. 
These (2,8) strings are left-moving fermions on the intersection  onebrane.
The (6,8) strings can also carry some charge but they 
are massive when the (2,6) strings are excited \dougauge .
As in the case of rotating black holes \spin\ ~we 
conclude\foot{Similar observations 
have been made by C. Vafa (private communication).} that the effective
momentum that is left  to  distribute in (2,6) strings, after we have
put
enough (2,8) strings to account for the charge, 
is 
$n_{eff} = n - q^2/2Q_1 $, where the factor of $Q_1$ arises as in 
\spin\ from 
the different flavors of (2,8) strings among which the 
charge is distributed. 
The entropy formula becomes 
\eqn\entroch{
S= 2 \pi \sqrt{ Q_1 n_{eff} Q_{KK} Q_5 } = 2 \pi \sqrt{ \biggl(
 Q_1 n -\half{ \bf
q}^2 \biggr) Q_{KK} Q_5 } = 2 \pi \sqrt{ {\bf  Q} ^2 {\bf P}^2 - 
( {\bf Q} \cdot {\bf P})^2},} 
since ${\bf Q}^2 = Q_1 n  -\half \bf q^2 $ and  in this case 
${\bf Q} \cdot {\bf P} =0 $. This is the classical formula
 \classfour. 
Here $\bf q^2$ is an even integer, each left-moving (2,8) fermion
carries one unit of charge and the 
total current carrying fermion number is restricted to even values \wittenpol.

It is also of interest to consider a black hole that is extremal (in
the sense that the mass is such that the solution is on the threshold 
of developing a naked singularity) but not 
BPS, as for example a black hole with ${ \bf Q}^2 <0 $. 
It can be seen from 
the general black hole solutions in \classfour ~that 
the classical entropy formula 
is just 
\eqn\nonextr{
S_{ext} = 2 \pi \sqrt{ | {\bf Q}^2 | {\bf P} ^2 }.
}
To be definite consider a black hole with zero momentum $n=0$, but
with
some gauge charge $\bf q$, so that  ${\bf Q} = (\half Q_1 R, -\half Q_1 R, 
\halfsq {\bf q})$, ${\bf Q}^2 = -\half {\bf q}^2$.
In this case, we have to have enough (2,8) strings to
carry the charge, but  since the $(2,8)$ strings can move only in one 
direction \wittenpol\ ~they will carry some net momentum along the
direction of the original string. This implies that there must be 
an equal amount
of open strings moving in the opposite direction. These will be
(2,6) strings since they 
carry the most entropy.  The black hole, and the D-brane system, 
are  not BPS but they are  extreme  
in the sense that they carry the minimum amount of mass consistent
with the given charges. 
Note that there is no BPS bound for the charges $\bf q$ under which
this black hole is charged. In fact there can be light particles 
charged under this gauge group near an enhanced symmetry point. 
A real-world electrically charged extremal black hole will
be of this type,  in the sense that the electron is nearly massless
compared to the string scale. 
Such black holes are not stable and will decay quickly by emitting
charged particles. 

\newsec{Entropy for more general $N=2$ cases}

Now we construct a black hole with charges that can exist
only in the $N =2$ case and not in the $N=4$ case. 
In order to do this we use the Gimon-Polchinski model \gimon\ which 
is connected to the Type I on K3 considered above.
This model contains 9-branes and 5-branes. The 5-branes 
are oriented along the directions (012345).
We compactify the directions (45) and T-dualize along 
the direction 4. Now we have 2 orientifold hyperplanes; 
the 9-branes (5-branes) transform into 16 8-branes
(4-branes) between the orientifold
hyperplanes. The 4-branes are oriented 
along (01235) and one can choose a point in moduli
space where there are not any coinciding branes. There are also 
orbifold fixed points on the internal torus (6789).
These branes are ``background'' branes in the sense that they
are completely extended along the macroscopic four dimensional
space and are part of the vacuum state.
 
Now we include the same configuration that we had before: 
$Q_5$ solitonic 5-branes along (056789), $Q_6$ 6-branes
along (0456789), $Q_2$ 2-branes along (045) and momentum $n$ 
along the direction 5. If these are all the charges we have,
the state counting for this case is the same as in
the previous case, since all ``background'' branes give contributions
that are subleading in the limit of large  charges.

The new feature, relative to the N=4 case, is that we can
have extra charges associated to 4-branes. 
These charges will be carried by (4,6) strings which are
left-moving fermions, they are related by T-duality to the
(2,8) or (1,9) strings of the previous section.
The (2,4) strings also carry charge, but become massive when
a condensate of (2,6) strings form. The (4,8) strings are a purely
subleading contribution since they involve only the background branes,
and, in any case, we can sit at a point in moduli space where they are
massive.

If the black hole also
carries charge $\bf p$ under the 4-brane $U(1)$s and
charge $\bf q$ under the 8-branes $U(1)$s,  then we are forced
to have some left-moving (2,8) and (4,6) strings thus
reducing the available momentum that we can distribute 
among the highly entropic (2,6) modes by $n_{eff}=
n - {\bf q}^2/2Q_2 - {\bf p}^2/2Q_6 $. The formula for
the entropy then becomes
\eqn\entrontwo{
S =  2 \pi \sqrt{ \biggl(Q_2 Q_6 n - {1\over 2} Q_2{\bf p}^2 - \half Q_6 {\bf
q}^2 \biggr) Q_5  }~.
}

\subsec{Classical Solution}

The Gimon-Polchinski model is U-dual to a heterotic theory on $K3$
with a instanton numbers (12,12) embedded in the two $E_8$ factors 
\refs{\seiwit, \berk}.
The six-dimensional low-energy lagrangian for this heterotic theory
has been considered in \dmw. This is equivalent to the
Type I action. The relevant terms in this action, in heterotic
variables,
are:
\eqn\lagsix{
\eqalign{
 S &= {(2\pi)^3 \over \alpha'^2 }
\int d^6 x \sqrt{-g} \biggl( e^{-\phi} \biggl[ R+ (\del \phi)^2 - 
{1\over 12} H^2- {\alpha'\over 8} F^2 \biggr]  -
{\alpha'\over 8} \tilde F^2 \biggr) \cr & +  
{(2\pi)^3 \over \alpha'^2 } \int_{M_6} -{\alpha' \over 4} 
B\wedge \tilde F \wedge \tilde F - {\alpha'^2 \over 8 } \omega_3 \wedge
\tilde \omega_3 ~,\cr}
}
where $F$ ($\tilde F$) denotes the field strength 
of the gauge fields arising from the 9-branes (5-branes), 
$\phi$ is the six-dimensional dilaton,
and $\omega_3$ is the Chern-Simons form defined by 
$d \omega_3= - F\wedge F/2 $, and likewise for $\tilde \omega_3$. 
The field strength for the antisymmetric  tensor field is defined
in the usual way as $H=dB+{\alpha' \over 2}  \omega_3$. Note
we have dropped the higher derivative terms that appear in the 
action of \dmw\ which will only be relevant for large curvatures.

The equations of motion that follow from this action are invariant
under a $\IZ_2$ duality transformation which acts as
\eqn\gpdual{
\eqalign{
\phi &\to - \phi \cr
g_{mn} &\to e^{-\phi} g_{mn} \cr
H &\to e^{-\phi} * H \cr
A &\to \tilde A \cr
\tilde A &\to A ~.\cr}
}
This symmetry is actually just a T-duality  symmetry 
on the Type I side  which
inverts the size of the $K3$.

Now let us compactify on a torus down to four dimensions and
consider the classical
black hole solution which
carries the charges mentioned above. It follows from \refs{\fks,\as}
that there exists a solution with constant scalar fields,
provided the asymptotic values of these scalars are adjusted
to special values. Because the entropy does not depend on the
asymptotic
 values
of the moduli \refs{\fks,\as} there is no loss of generality in restricting
our considerations to this case. For this solution, it may be seen from the
equations of motion that $H= *H$, where the Hodge dual $*$ is defined
with respect to string metric. Taking into account the
fact that $F\wedge F$ and $\tilde F \wedge \tilde F$ vanish
for the solution at hand, the equations of motion take the same
form as the usual $N=4$ Type I (or heterotic) equations \maharana.
The extremal BPS black hole solutions of the $N=4$ equations have
been classified in \classfour. Using these results it
may be shown that the field $\phi$ satisfies
\eqn\phiis{
e^{\phi} = {Q_6 \over Q_2}~,
}
and the entropy is
\eqn\clentrop{
S= 2\pi \sqrt{\biggl(Q_2 Q_6 n - \half Q_6 
({\bf q}^2 + {\bf s}^2) \biggr)Q_5}~,
}
where all charges are as defined above, and $\bf s= \gamma \bf p$ with
$\gamma$ a constant to be determined.

The difference between the $F$ and $\tilde F$ fields
arises when one when considers the relationship
between the integer-valued quantized charges and the physical charges $Q^i$
(defined by ${\cal F}^i = Q^i/r^2$, with
${\cal F}^i$ the four-dimensional field strengths, as defined in
\maharana). Since the gauge kinetic terms for the $\tilde F$ fields
do not have the usual $e^{-\phi}$ factor in front, we find the
relation ${\bf s} = e^{-\phi/2} \bf p$. Substituting this into
\phiis\ and \clentrop\ we find the Bekenstein-Hawking entropy of the
black hole agrees with the microscopic counting \entrontwo.

\newsec{ Conclusions }

We have found agreement between the macroscopic Bekenstein-Hawking
entropies for BPS black holes in $N=2$ supergravities
 and the microscopic entropy in string theory. 
In the first two examples the counting is very similar to 
the counting for the $N=4,8$ cases. The only real 
difference is that
the various branes are on a less supersymmetric background. The
physical
mechanism that gives rise to the large degeneracy is basically the
same as in the more symmetric cases.  
We explored more intrinsically $N=4,2$ cases by considering black 
holes which carry gauge charges that exist in these less
supersymmetric
theories.

It would be interesting to present a general argument testing the
full $N=2$ spectrum of
charged black holes. In particular,  D-brane counting 
for Type II string theories compactified   
on generic Calabi-Yau 3-folds that are not orbifolds of more symmetric cases
is an unexplored problem. The results of \refs{\as, \fk} describe a 
universal geometric formula for the entropy which must be somehow
reproduced by D-branes. In particular the simple relationship \fk\ of 
the entropy formula to the minima of the central charge should 
have a microscopic explanation.

\bigskip
{\bf Acknowledgements}

We thank J. Park and J.H. Schwarz for helpful discussions.
The research of D.L. is supported in part by
DOE grant DE-FG03-92ER40701; that of 
J. M. by DOE grant DE-FG02-96ER40559; and that of
D. K. and A. S. by DOE grant DOE-91ER40618.

\listrefs
\end